\documentclass[letter]{aa} 

\usepackage{graphicx}

\usepackage{txfonts}
\newcommand{\nhtoz}{NH$_3$(1$_{0}$-0$_{0}$)}
\newcommand{\nht}{NH$_3$}

\newcommand{\htooo}{H$_2$O(1$_{10}$-1$_{01}$)}
\newcommand{\hto}{H$_2$O}
\newcommand{\htwo}{H$_2$}

\newcommand{\ceo}{C$^{18}$O(1-0)}
\newcommand{\htdp}{H$_2$D$^+$(1$_{10}$-1$_{11}$)}
\newcommand{\nthp}{N$_2$H$^+$(1-0)}
\newcommand{\hcop}{HCO$^+$(1-0)}
\newcommand{\percc}{cm$^{-3}$}
\newcommand{\kms}{km\,s$^{-1}$}

\newcommand{\vlsr}{V$_{\rm LSR}$}
\begin{document}

   \title{NH$_3$ (1$_{0}$-0$_{0}$) in the pre-stellar core L1544\thanks{Based on observations carried out with {\em Herschel}, an ESA space observatory with science instruments provided by a European-led Principal Investigator consortium
and with important participation from NASA.}}


   \author{P. Caselli
          \inst{1,2}
          \and
          L. Bizzocchi\inst{1}
          \and
          E. Keto\inst{3}
          \and 
          O. Sipil{\"a}\inst{1}
          \and 
          M. Tafalla\inst{4} 
          \and
          L. Pagani\inst{5}
          \and
          L. E. Kristensen\inst{6}
          \and
          F. F. S. van der Tak\inst{7,8}
          \and
          C. M. Walmsley\inst{2,9}\thanks{Deceased.}
          \and
          C. Codella\inst{2}
          \and
        B. Nisini\inst{10}
          \and
          Y. Aikawa \inst{11}
          \and
          A. Faure\inst{12}
          \and
          E. F. van Dishoeck\inst{13,1}
          }

   \institute{Centre for Astrochemical Studies, 
              Max-Planck-Institute for Extraterrestrial Physics, Giessenbachstrasse 1, 85748 Garching, Germany; 
              \email{caselli@mpe.mpg.de}
         \and
         	INAF-Osservatorio Astrofisico di Arcetri, Largo E. Fermi 5, 50125 Firenze, Italy\\
         \and
             Harvard-Smithsonian Center for Astrophysics, 60 Garden Street, Cambridge,  MA 02138, USA\\
          \and
          	Observatorio Astron{\'o}mico Nacional (IGN), Calle Alfonso XII, 3, Madrid, Spain\\
	  \and
		LERMA and UMR 8112 du CNRS, Observatoire de Paris, 61 Av. de l'Observatoire, 75014 Paris, France\\
	  \and
	  	Centre for Star and Planet Formation, Niels Bohr Institute and Natural History Museum of Denmark, University of Copenhagen, 
		{\O}ster Voldgade 5-7, 1350, Copenhagen, Denmark\\
	\and
		SRON Netherlands Institute for Space Research, Landleven 12, 9747 AD, Groningen, The Netherlands\\
	\and
		Kapteyn Astronomical Institute, University of Groningen, The Netherlands\\
	\and
		The Dublin Institute for Advanced Studies, 31 Fitzwilliam Place, Dublin 2, Republic of Ireland\\
	\and
		INAF - Osservatorio Astronomico di Roma, 00040 Monte Porzio Catone, Italy\\
	\and
		Center for Computational Science, University of Tsukuba, Tsukuba, Ibaraki 305-8577, Japan\\
	\and
		Universit\'e Grenoble Alpes, CNRS, IPAG, F-38000 Grenoble, France\\
	\and 
		Leiden Observatory, Leiden University, PO Box 9513, 2300 RA Leiden, The Netherlands\\
             }

   \date{Received December 15, 2016; accepted March 16, 2017}

 
  \abstract{Pre-stellar cores represent the initial conditions in the process of star and planet formation, therefore it is important to study their physical and chemical structure. Because of their volatility, nitrogen-bearing molecules are key to study the dense and cold gas present in pre-stellar cores.  The NH$_3$ rotational transition detected with {\em Herschel}-HIFI provides a unique combination of sensitivity and spectral resolution to further investigate physical and chemical processes in pre-stellar cores.  Here we present the velocity-resolved {\em Herschel}-HIFI observations of the ortho-\nhtoz\ line at 572\,GHz and study the abundance profile of ammonia across the pre-stellar core L1544 to test current theories of its physical and chemical structure. 
  Recently calculated collisional coefficients have been included in our non-LTE radiative transfer code to reproduce {\em Herschel} observations. A gas-grain chemical model, including spin-state chemistry and applied to the (static) physical structure of L1544 is also used to infer the abundance profile of ortho-\nht . The hyperfine structure of ortho-\nhtoz\ is resolved for the first time in space. All the hyperfine components are strongly self-absorbed.  The profile can be reproduced if the core is contracting in quasi-equilibrium, consistent with previous work, and if the \nht\ abundance is slightly rising toward the core centre, as deduced from previous interferometric observations of para-\nht (1,1). The chemical model overestimates the NH$_3$ abundance at radii between $\simeq$ 4000 and 15000\,AU by about two orders of magnitude and underestimates the abundance toward the core centre by more than one order of magnitude.  Our observations show that chemical models applied to static clouds have problems in reproducing NH$_3$ observations.}


   \maketitle
%

\section{Introduction}

Since its discovery as the first polyatomic molecule in space \citep{Cheung1968}, ammonia has been widely used as a temperature and structural probe of dense cloud cores in low-mass \citep[e.g.][]{Benson1989} and high-mass \citep[e.g.][]{Pillai2006} star-forming regions, and as a tracer of shocks along outflows driven by young stellar objects \citep[e.g.][]{Tafalla1995}.  NH$_3$ can form on the surface of dust grains through hydrogenation of atomic nitrogen and in the gas phase soon after the formation of N$_2$.  Unlike molecules such as CO, NH$_3$ does not appear to be affected by freeze-out within dense and cold starless cores \citep{Tafalla2002}, despite its high binding energies (close to those of H$_2$O). While the inversion transitions of ortho (o) and para (p) \nht\, around 1.3\,cm are easily accessible from the ground, the rotational transitions fall into the sub-millimeter and far-IR, and these can only be detected using space-borne receivers. The Einstein coefficients of the rotational transitions of NH$_3$ are about 10,000 times larger than those of the inversion transitions \citep{Ho1983},  implying significantly higher critical densities \citep[$\simeq$10$^7$\,\percc\ vs. $\simeq$10$^3$\,\percc ;][]{Danby1988}.   

The first observation of the ground-state rotational transition of o-\nhtoz\, was carried out by \citet{Keene1983} toward OMC-1 using the Kuiper Airborne Observatory (KAO). About twenty years later, the Odin satellite detected o-\nhtoz\, toward the Orion Bar \citep{Larsson2003}, high-mass star-forming regions \citep{Hjalmarson2005, Persson2007, Persson2009} and, for the first time, toward a dark cloud \citep[$\rho$\,Oph A;][]{Liseau2003}. Multiple rotational transitions have also been detected using the Heterodyne Instrument for the Far Infrared \citep[HIFI;][]{deGraauw2010} instrument on board the {\em Herschel Space Observatory} in the direction of high-mass star forming regions \citep{Persson2010,  Gerin2016} and with the {\em Infrared Space Observatory} \citep[ISO;][]{Ceccarelli2002}. The  o-\nhtoz\, has also been detected with {\em Herschel} by \citet{Codella2010} toward the shock region L1157-B1, by \citet{Salinas2016} toward the protoplanetary disk TW Hydrae, and by \citet{Lis2016} in the direction of the starless core L1689N, next to the young protostar IRAS16293-2422. None of the above spectra show resolved hyperfine structure of o-\nht, including the one toward L1689N, where only one group of hyperfine components is detected.  The 3$_{2}$-2$_{2}$ rotational transition of \nht\, at 1.8\,THz has also been observed with the Stratospheric Observatory For Infrared Astronomy (SOFIA) by \citet{Wyrowski2012, Wyrowski2016} to measure the infall rate in high-mass star-forming clumps.  

We report the first detection of the hyperfine structure of o-\nhtoz\, toward the isolated pre-stellar core L1544 in the Taurus molecular cloud complex, as part of the Water In Star-forming regions with {\em Herschel} (WISH) key project \citep{vanDishoeck2011}. The turbulence in L1544 is subsonic and the narrow spectral line widths, together with the high spectral resolution of HIFI, allow us to distinguish the quadrupole hyperfine structure that is due to the interaction between the molecular electric field gradient and the electric quadrupole moment of the nitrogen nucleus. These observations also confirm the dynamical structure that has been deduced by previous theoretical studies \citep{Keto2008, Keto2010, Keto2015} constrained by ground-based \citep{Caselli1999, Caselli2002} and {\em Herschel} \citep{Caselli2010, Caselli2012} observations. In the following, we refer to the core envelope as the region beyond $\simeq$4000\,AU, where the number density drops below 10$^5$\,\percc\ \citep{Keto2010}. Observations are described in Sect.\,\ref{sec_obs}, the o-\nhtoz\, spectrum is shown in Sect.\,\ref{sec_res} and the analysis is presented in Sect.\,\ref{sec_ana}. Discussion and conclusions can be found in Sect.\,\ref{sec_dis}. 

\section{Observations} \label{sec_obs}
The o-\nhtoz \ line \citep[central frequency 572.49815977\,GHz $\pm$0.2\,kHz;][]{Cazzoli2009} has been observed with {\em Herschel} toward the pre-stellar core L1544 (coordinates RA(J2000) = 05$^h$04$^m$17$^s$.21, Dec(J2000) = 25$^{\circ}$10\arcmin42\arcsec .8) for 10.3\,hours on April 2, 2011, together with the o-\htooo \ \citep{Caselli2012}. The data presented here are archived at the {\em Herschel} Science Archive\footnote{\url{http://archives.esac.esa.int/hda/ui}} with the identification number (OBSID) 1342217688. The technical details of the observations, which include the simultaneous use of the wide-band (WBS) and the high-resolution (HRS) spectrometers of HIFI in both H and V polarisation, have been described in \citet{Caselli2012}. We briefly summarise the details related to the HRS data for o-\nht , which are used here: the half-power beam width (HPBW) of the telescope at 572\,GHz is 40\arcsec \ and the spectral (nominal) resolution is 64 m\,s$^{-1}$ at 572.5\,GHz.  The individual spectra were reduced using the {\em Herschel} Interactive Processing Environment \citep[HIPE;][]{Ott2010}, version 14.1, and subsequent analysis of the data was performed with the Continuum and Line-Analysis Single dish Software (CLASS) within the GILDAS package\footnote{\url{http://www.iram.fr/IRAMFR/GILDAS}}. The latest in-flight-measured beam efficiencies were used to convert the intensity scale from antenna temperature into main-beam temperature (T$_{\rm mb}$)\footnote{\url{http://herschel.esac.esa.int/twiki/bin/view/Public/HifiCalibrationWeb}}. The H and V polarisation spectra are identical and they have been summed together to increase the sensitivity. 

 \section{Results} \label{sec_res}
   \begin{figure}
   \includegraphics[width=9cm]{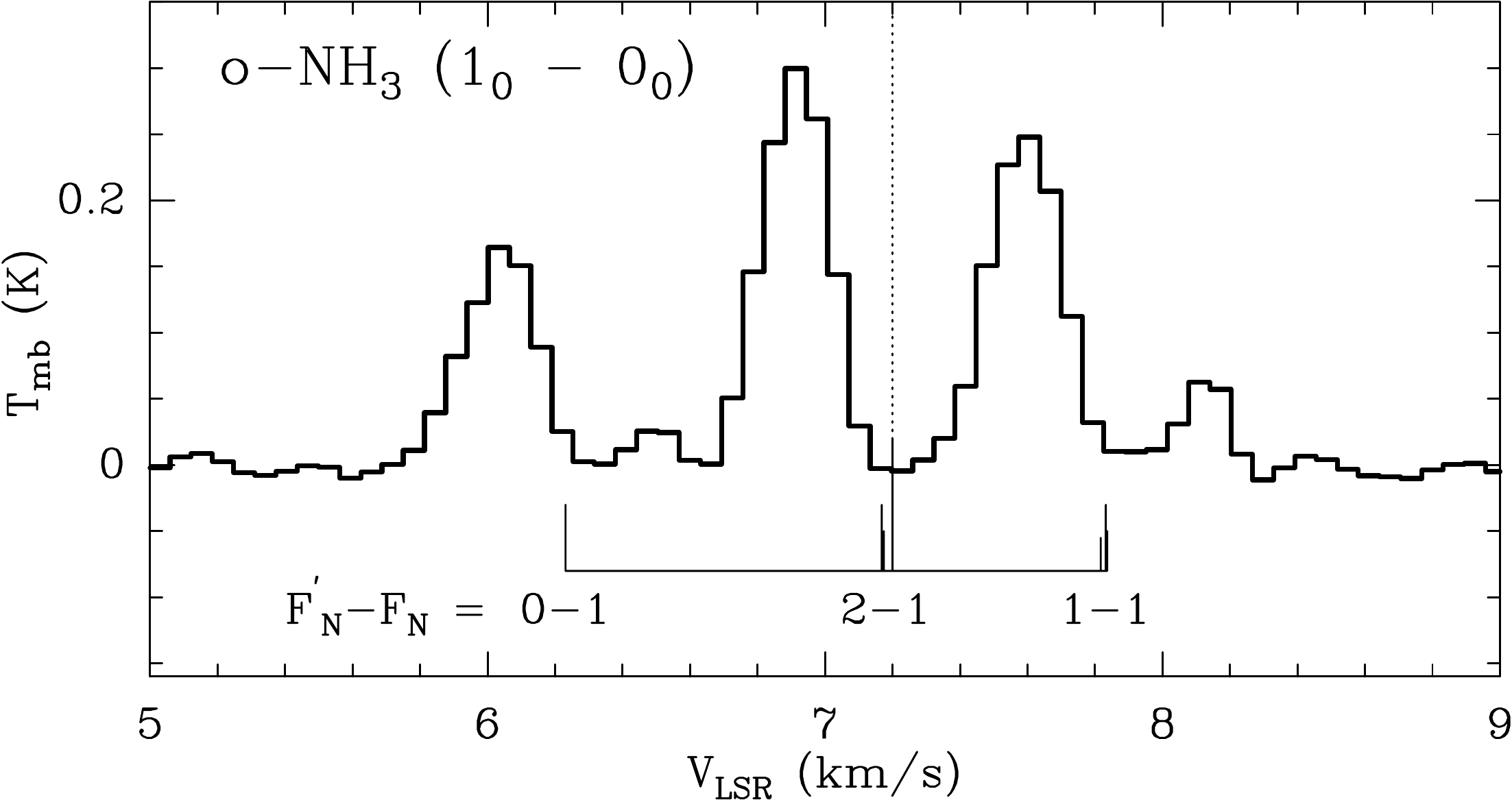}
   \caption{{\em Herschel} HIFI spectrum of o-\nhtoz\, toward the pre-stellar core L1544. The three groups of hyperfine components (with the 0-1 group having one single component) are detected for the first time in space.  They are all highly self-absorbed (see Section\,\ref{sec_res}). The hyperfine structure is shown by black vertical lines (with LTE relative intensities) and labelled as in  \citet{Cazzoli2009}. The vertical dotted line marks the core velocity. The root-mean-square noise level is 4.7\,mK.}
              \label{FigSpectrum}%
    \end{figure}

The continuum-subtracted o-\nhtoz, centred at the frequency of the main hyperfine component (572.4983387\,GHz), is shown in Fig.\ref{FigSpectrum} together with the hyperfine structure from Table\,3 of \citet{Cazzoli2009}, adopting the Local Standard of Rest (LSR) velocity (\vlsr) of L1544, 7.2\,\kms, as reported by \citet{Tafalla1998}. This is the first observed interstellar o-\nhtoz \ spectrum clearly showing the main groups of hyperfine components. The mismatch between the  \vlsr\ and the centroid velocity of the main group of hyperfine components is due to heavy self-absorption (from the envelope) combined with Doppler-shifted emission \citep[due to core contraction;][]{Keto2010}. In fact, \nht\ is abundant across L1544 \citep{Crapsi2007}, and the outer portions of the contracting pre-stellar core, where the volume density is significantly lower than the critical density of o-\nhtoz , are absorbing the redshifted part of the spectrum that is emitted toward the centre.

\begin{figure*}
   \centering
   \includegraphics[width=\hsize]{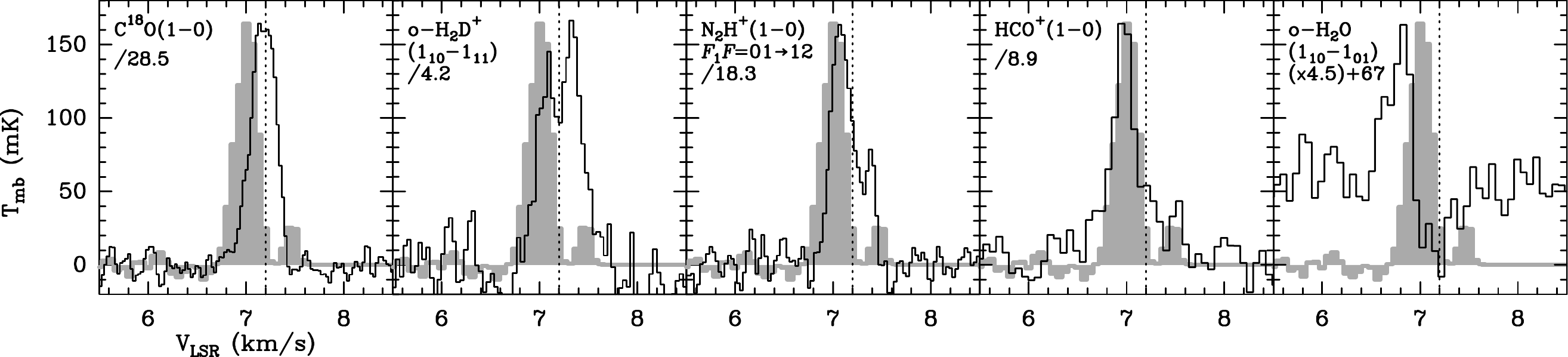}
   \caption{ 
Comparison between the isolated hyperfine component  $F^{\prime}_{\rm N}$ $\rightarrow$ $F_{\rm N}$ = 0\,$\rightarrow$\,1 of o-\nhtoz \ (filled grey histogram; for clarity, the NH$_3$ spectrum has been cut above 7.8\,\kms (6.6\,\kms \ in the velocity scale of Fig.\,\ref{FigSpectrum}), where the other groups of hyperfine components are present and other molecular lines observed toward the same position: \ceo , o-\htdp , \nthp , \hcop , and o-\htooo \, (thin histograms). To allow a clearer profile comparison, the   \ceo , o-\htdp , \nthp , and \hcop \ spectra have been divided by the numbers shown below the transition in the top left corner of the panels. In the case of o-\htooo \, (last panel on the right), the spectrum has first been multiplied by 4.5 and then shifted by $+$67\,mK. The comparison spectra are in order of increasing optical depth, from left to right. The vertical dotted line marks the core velocity.
   }
              \label{Fig_Comparison}%
    \end{figure*}

Self-absorption and Doppler-shifting due to the dynamics have already been detected in other molecular lines when observed toward the same line of sight.  Figure\,\ref{Fig_Comparison} shows a comparison of the only isolated hyperfine component in the spectrum of o-\nhtoz \ ($F^{\prime}_{\rm N}$ $\rightarrow$ $F_{\rm N}$ = 0\,$\rightarrow$\,1; filled grey histogram) with other lines: \ceo , o-\htdp , \nthp , \hcop , and o-\htooo\ (black histograms).  The o-NH$_3$ hyperfine component displays the characteristic blue-shifted infall profile \citep[e.g.][]{Evans1999}, with a self-absorption dip reaching the zero-level, thus indicating a large line optical depth \citep{Myers1996}. It is interesting to note that the NH$_3$ line shows extra emission at low velocities (coming from the material approaching the core centre from the back) when compared to the low-density tracer \ceo , which does not probe the central regions because of the large CO freeze-out \citep{Caselli1999}. Extra blue emission of the o-\nhtoz \ hyperfine component is also seen when compared to the o-\htdp \, and  \nthp \ lines, although to a lesser extent. The 'blue excess' disappears in the case of \hcop , which is also heavily absorbed \citep{Tafalla1998}. The resemblance between o-\nhtoz \, and \hcop \ is quite striking, in view of the fact that the distribution of HCO$^+$ across the core is known to be quite different from that of \nht , with HCO$^+$ tracing the outer layers of the core while being depleted toward the centre because of the freeze-out of its main parent species CO \citep{Tafalla2002,Tafalla2006}. The H$_2$O line is the most blue-shifted and displays the broadest absorption, thus suggesting higher optical depths \citep[see][]{Keto2014}. This is probably caused by the large abundance of H$_2$O (compared to NH$_3$) in the L1544 outer layers, which is responsible for the absorption in a velocity range spanned by the lower density tracers in Fig.\,\ref{Fig_Comparison}. 
 
 \section{Analysis} \label{sec_ana}
  
The non-Local Thermal Equilibrium (non-LTE) radiative transfer code MOLLIE \citep{Keto1990, Keto2004, Keto2010b} was used to interpret the o-\nhtoz \ spectrum 
in Fig.\,\ref{FigSpectrum}. We took into account the hyperfine structure, produced by the electric quadrupole coupling of the N nucleus with the electric field of the electrons, as well as the magnetic hyperfine structure that is due to the three protons \citep{Cazzoli2009}. The hyperfine collision rate coefficients were taken from the recent calculations of Bouhafs et al. (in preparation), which, for the first time, include the non-spherical structure of p-H$_2$ \citep[the main form of H$_2$ in cold molecular gas,][]{Flower2006,Brunken2014}, so that they differ (by a factor larger than $\simeq$2) from those of \citet{Maret2009}. The calculations were restricted to the lowest nine hyperfine levels of o-NH$_3$, corresponding to the first two rotational levels (0$_0$ and 1$_0$), within a temperature range 5-30 K. Since only the ground-state o-NH$_3$ transitions were considered, no scaling of the rotational rates was necessary: all hyperfine rates are equal to the pure rotational rates, and intra-multiplet rates are set to zero, as in the standard statistical approach.  

\begin{figure}
   \centering
    \includegraphics[width=9cm]{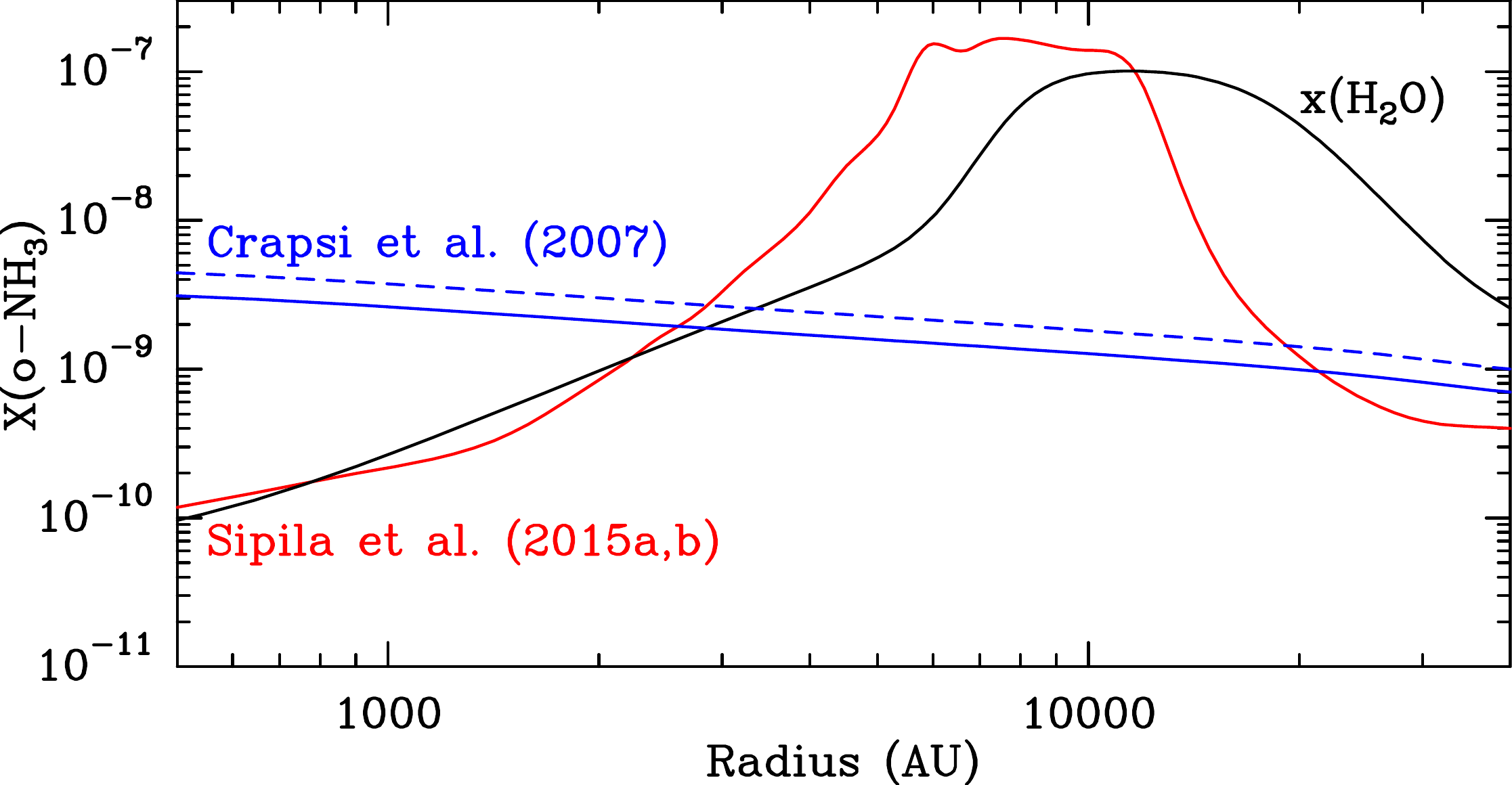}
      \caption{Radial profile of the fractional abundance of o-\nht\,  with respect to \htwo , X(o-\nht), in L1544. The blue profile has been deduced from observations carried out by \citet{Crapsi2007} assuming an ortho-to-para NH$_3$ abundance ratio of 1 (dashed curve) and 0.7 (solid curve); the red profile is the result of a chemical model calculation by \citet{Sipila2015a,Sipila2015}. The black profile refers to the total (ortho+para) abundance of H$_2$O from \citet{Keto2014}. 
              }
         \label{FigXNH3}
   \end{figure}

Unlike for H$_2$O and CO \citep{Keto2008, Caselli2012, Keto2014}, we did not develop any simplified chemistry for NH$_3$ to be coupled with the hydro-dynamical evolution. Two different approaches were followed to simulate the observed spectrum: (1) the radial profile of o-\nht\ follows that of p-\nht\ deduced by \citet{Crapsi2007} using VLA observations of p-\nht (1,1), assuming an ortho-to-para NH$_3$ abundance ratio of 1 and 0.7\footnote{Taking into account the calibration errors of the \citet{Crapsi2007} observations, we cannot distinguish between these two values. However, the two slightly different abundance profiles show how possible small variations in the abundance can affect the simulated spectrum.}, and extrapolating to larger radii (see Eq.\,\ref{EqProfile}); 
(2) the radial profile of o-\nht\ was calculated using the chemical model predictions of \citet{Sipila2015a,Sipila2015} and \citet{Sipila2016} applied to the L1544 physical structure deduced by \citet{Keto2010} and \citet{Keto2015}, who demonstrated that other observed line profiles can be reproduced by the quasi-equilibrium contraction of an unstable Bonnor-Ebert sphere.  The two abundance profiles are presented in Figure\,\ref{FigXNH3} together with the profile of H$_2$O deduced by \citet{Caselli2012} and refined by \citet{Keto2014}.

\citet{Crapsi2007} followed the same analysis as \citet{Tafalla2002}. They parametrised the abundance profile of total (ortho+para) \nht , X$_{\rm NH_3}$($r$), starting from X(p-NH$_3$) and assuming an ortho-to-para ratio of unity, with the function: 
\begin{equation}
{\rm X_{NH_3}} (r)  =  {\rm X_0} \left( \frac{n(r)}{n_0} \right) ^{{\alpha}_{\rm X}} ,
\label{EqProfile}
\end{equation}
and then finding the best-fit parameters: X$_0$ =  8$\times$10$^{-9}$, $n_0$ = 2.1$\times$10$^{6}$\,\percc \, (the average density within a radius of 14${\arcsec}$) and ${\alpha}_{\rm X}$ = 0.16.  The abundance profile of p-NH$_3$ measured by \citet{Crapsi2007} is given by multiplying the right-hand side of Eq.\,\ref{EqProfile} by 0.5; from this, we derived the abundance profile of o-NH$_3$ assuming an ortho-to-para NH$_3$ ratio of 1 (dashed blue curve in Fig.\,\ref{FigXNH3}) and 0.7 (solid blue curve in Fig.\,\ref{FigXNH3}). The latter value is expected in cold gas and would indicate a gas-phase formation for NH$_3$ \citep{Faure2013}. As noted by \citet{Tafalla2002}, \citet{Crapsi2007} found that the NH$_3$ abundance slightly increases toward the core centre. We note that the interferometric observations of \citet{Crapsi2007} have an angular resolution of about 4\arcsec \ ($\sim$300\,AU in radius), so they place stringent constraints on the abundance profile of NH$_3$\footnote{The hyperfine structure of p-NH$_3$(1,1) allows for abundance measurements even at large optical depths.}. A comparison of the blue curves with the (ortho+para) H$_2$O abundance profile deduced by \citet{Keto2014} (black curve in Fig.\,\ref{FigXNH3}) shows that the total NH$_3$ is about 30 times more abundant than H$_2$O within the central 1000\,AU, while it is two orders of magnitude less abundant than H$_2$O from about 8000\,AU to 20000\,AU.  

The o-NH$_3$ profile has a very different shape when the chemical model of \citet{Sipila2015a,Sipila2015}, which includes spin-state chemistry, is considered (red curve in Fig.\,\ref{FigXNH3}).  
The chemical abundance gradients in L1544 were simulated using the same procedure as described in \citet{Sipila2016}: the physical model for L1544 \citep{Keto2010,Keto2014} was separated into concentric shells and the chemical evolution was calculated separately in each shell. The gas was initially atomic with the exception of H (all in H$_2$) and D (in HD). The o-NH$_3$ abundance profile (Fig.\,\ref{FigXNH3}) was extracted from the model at the time when the CO column density was comparable to the observed value \citep[$1.3 \times 10^{18} \, \rm cm^{-2}$; see][]{Caselli1999}, taking into account telescope beam convolution. Other details of the chemical model are extensively discussed in \citet{Sipila2015a,Sipila2015}. The predicted o-NH$_3$ profile (red curve in Fig.\,\ref{FigXNH3}) presents a broad peak between 5000\,AU and 15000\,AU, it drops at larger radii because of photodissociation and at lower radii because of freeze-out. 
   
\begin{figure}
   \centering
    \includegraphics[width=9cm]{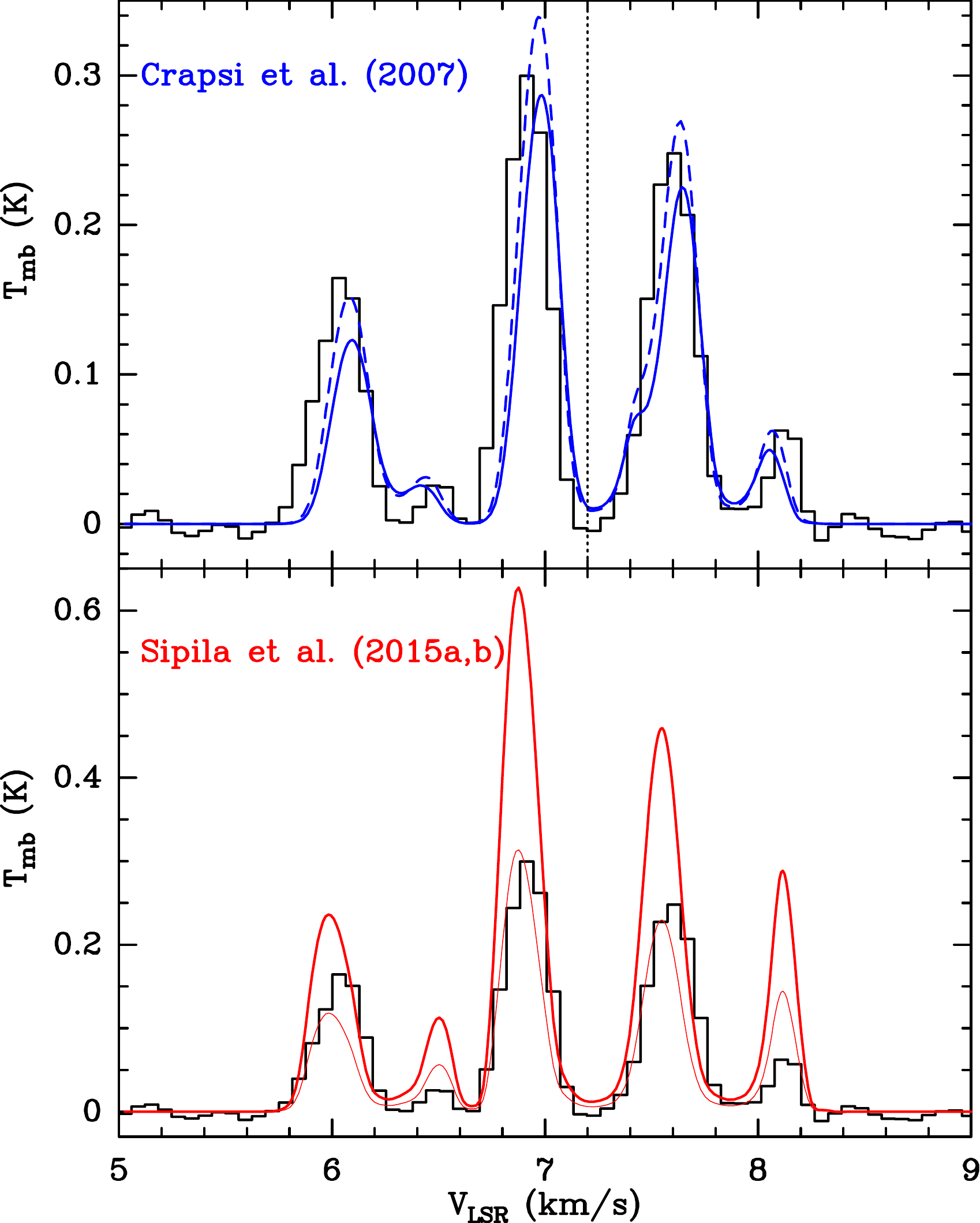}
      \caption{MOLLIE radiative transfer results using the o-NH$_3$ fractional abundance profiles from Fig.\,\ref{FigXNH3}, overlapped with the observed spectrum: ({\it top}) blue solid and dashed curves  refer to the \citet{Crapsi2007} profiles, assuming an ortho-to-para NH$_3$ ratio of 0.7 and 1.0, respectively. ({\it Bottom}) The red thick curve refers to the \citet{Sipila2015a,Sipila2015} profile; the red thin curve is the same as the thick red curve, but divided by 2. 
              }
         \label{Fig_RadTrans}
   \end{figure}

 The o-\nhtoz\ spectra simulated with MOLLIE are shown in Fig.\,\ref{Fig_RadTrans} for the three o-NH$_3$ abundance profiles in Fig.\,\ref{FigXNH3}, together with the observed spectrum. This figure shows that the empirical abundance profile of \citet{Crapsi2007} reproduces the observed spectrum better than the abundance profile predicted by \citet{Sipila2015a,Sipila2015}, although the simulated spectrum underestimates the brightness temperature at the lowest velocities of all groups of hyperfine components, as also found in the case of o-H$_2$O \citep{Caselli2012}, which suggests that the contraction velocity within the core centre is slightly higher than the velocity used in our model.  The red spectrum \citep[based on the chemical model of][]{Sipila2015a,Sipila2015} is significantly brighter than the detected line and the ratios between the low- and high-velocity peaks of the various hyperfine components are not well reproduced (as better shown by the red thin curve in Fig.\,\ref{Fig_RadTrans}, which is the red thick profile scaled down by a factor of two). It is interesting to note that the red curve is shifted to lower velocities than the blue curves. This is due to the large abundance of \nht \ predicted at radii in the range of 4000-12000\,au (Fig.\,\ref{FigXNH3}, red curve), which produces a bright optically thick line and causes absorption around the core LSR velocity, as the contraction velocity is lower in the core envelope, which produces the absorption \citep{Keto2010}.  

 \section{Discussion and conclusions} \label{sec_dis}
  
{\em Herschel} observations of the o-\nhtoz \ line toward the dust peak of the L1544 pre-stellar core have been presented. The spectrum shows for the first time the three well-separated groups of hyperfine components. The three components are strongly self-absorbed and display the characteristic infall profile, with a blue peak significantly brighter than the red peak and a dip reaching the zero level, although for the $F^{\prime}_{\rm N}$ $\rightarrow$ $F_{\rm N}$ = 2\,$\rightarrow$\,1 components, the red peak is blended with the blue peak of the 1-1 component. The observed spectrum can be well reproduced adopting the empirical NH$_3$ abundance profile deduced by the interferometric observations of \citet{Crapsi2007} as well as  the physical structure that was used to reproduce previous spectra: the Bonnor-Ebert sphere in quasi-static contraction, which is characterised by a central density of 10$^7$\,\percc \ and a subsonic contraction velocity field peaking at a radius of 1000\,AU \citep{Keto2010, Caselli2012, Keto2015}. 

A poorer match with observations is seen when the o-\nht\ profile predicted by the comprehensive chemical model of \citet{Sipila2015a,Sipila2015}, applied to the (static) physical structure of L1544, is used as input in MOLLIE.  This suggests that 'pseudo-time dependent' chemical models (i.e. time-dependent chemical models applied to static clouds) have problems in reproducing the observed chemical structure of pre-stellar cores. The model predicts too much NH$_3$ in the outer parts of the cloud\footnote{Between 4000 and 15000\,AU.  Above 20000\,AU, the \nht\ abundance from \citet{Crapsi2007} is larger than that predicted by \citet{Sipila2015a,Sipila2015}; however, this does not affect the model profile based on Eq.\,\ref{EqProfile} in Fig.\,\ref{Fig_RadTrans}, as at radii $>$20000\,AU, the volume density is lower than 3000\,\percc , and the absorbing column is negligible.} and too little toward the centre, when compared with the abundance deduced by the interferometric observations of \citet{Crapsi2007}.  The predicted large NH$_3$ abundance at radii around 10000\,AU is mainly caused by surface formation followed by reactive desorption (with 1\% efficiency, which may be overestimated).  Figure\,\ref{Fig_Comparison} shows that the o-H$_2$O spectrum is more absorbed around the LSR velocity than that of o-NH$_3$, suggesting that H$_2$O is indeed more extended than NH$_3$ in the outer low-density part of the envelope, while the red curve in Fig.\,\ref{FigXNH3} predicts \nht\ abundances around 10000\,AU comparable to those of \hto \, (see Fig.\,\ref{FigXNH3}).  Toward the core centre, the predicted low NH$_3$ abundance indicates that selective desorption mechanisms for NH$_3$ \citep[e.g. lower binding energies for NH$_3$ adsorbed on CO and/or N$_2$ ice layers, formed on top of the water ice, combined with ice heating or disruptive action by cosmic rays;][]{Ivlev2015}, not included in the model, may be at work. Another possibility could be that the production of gas phase NH$_3$ is underestimated;  the N$_2$ abundance profile drops more slowly than the CO abundance profile in the other pre-stellar core L183 \citep{Pagani2012}, meaning that N$_2$ desorption is more efficient than currently advocated and/or that production of N$_2$ is ongoing in the gas phase, thus maintaining efficient production of NH$_3$ through the usual routes in the gas phase. However, the assumption of a static cloud could also be the cause of the discrepancy with the empirical profile, and this will be discussed in a future paper (Sipil\"a et al., in prep.). It is clear that more work needs to be done to gain understanding of the gas-grain chemical processes regulating \nht\  and N-bearing molecules in cold and quiescent objects such as L1544, the precursors of future stellar systems.

%

  \begin{acknowledgements}
P.C. acknowledges the financial support from the European Research Council (ERC; project PALs 320620). MT acknowledges
partial support from MINECO project AYA2016-79006-P. CMW acknowledges
 support from Science Foundation Ireland Grant 13/ERC/I2907. HIFI has been designed and built by a consortium of institutes and university departments from across Europe, Canada and the United States under the leadership of SRON Netherlands Institute for Space Research, Groningen, The Netherlands and with major contributions from Germany, France and the US. Consortium members are: Canada: CSA, U.Waterloo; France: CESR, LAB, LERMA, IRAM; Germany: KOSMA, MPIfR, MPS; Ireland, NUI Maynooth; Italy: ASI, IFSI-INAF, Osservatorio Astrofisico di Arcetri-INAF; Netherlands: SRON, TUD; Poland: CAMK, CBK; Spain: Observatorio Astron{\'o}mico Nacional (IGN), Centro de Astrobiolog{\'i}a (CSIC-INTA). Sweden: Chalmers University of Technology - MC2, RSS \& GARD; Onsala Space Observatory; Swedish National Space Board, Stockholm University - Stockholm Observatory; Switzerland: ETH Zurich, FHNW; USA: Caltech, JPL, NHSC.
 \end{acknowledgements}

%
\bibliographystyle{aa} 
\bibliography{nh3_references} 
%

\end{document}